\documentclass[useAMS]{mn2e}
\usepackage{psfig}

\usepackage{times}
\usepackage{amssymb,amsmath}

\def\lsim{ \lower .75ex\hbox{$\sim$} \llap{\raise .27ex \hbox{$<$}} }
\def\gsim{ \lower .75ex \hbox{$\sim$} \llap{\raise .27ex \hbox{$>$}} }

\title[Spine-layer model of NGC 1275] 
{On the spine-layer scenario for the very high-energy emission of NGC 1275}

\author[Tavecchio \& Ghisellini]
{F. Tavecchio$^1$\thanks{E--mail: fabrizio.tavecchio@brera.inaf.it} and
G. Ghisellini$^1$\\
$^1$INAF -- Osservatorio Astronomico di Brera, via E. Bianchi 46, I--23807
Merate, Italy\\
}

\begin{document}



\maketitle

\begin{abstract} 
We discuss the $\gamma$--ray emission of the radiogalaxy NGC 1275 (the central galaxy of the 
Perseus Cluster), detected by {\it Fermi}--LAT and MAGIC, in the framework of the ``spine-layer" 
scenario, in which the jet  is assumed to be characterized by a velocity structure, 
with a fast spine surrounded by a slower layer. 
The existence of such a structure in the parsec scale jet of NGC 1275 has been 
recently proved through VLBI observations.
We discuss the constraints that the observed spectral energy distribution 
imposes to the 
parameters and we present three alternative models, corresponding to three different choices of 
the angles between the jet and the line of sight ($\theta_{\rm v}=6^{\circ}, 18^{\circ}$ and 25$^{\circ}$). 
While for the the case with $\theta_{\rm v}=6^{\circ}$ we obtain an excellent fit, we consider 
this solution unlikely, since such small angles seems to be excluded by radio observations 
of the large-scale jet.  For $\theta_{\rm v}=25^{\circ}$ the required large intrinsic 
luminosity of the soft (IR--optical) component of the spine determines a large optical 
depth for $\gamma$--rays through the pair production scattering $\gamma \gamma\rightarrow e^+ e^-$, 
implying a narrow cut--off at  $\sim50$~GeV. 
We conclude that intermediate angles are required.
In this case the low frequency and the high--energy emissions are produced by 
two separate regions and, in principle, a full variety of correlations is expected. 
The correlation observed between the optical and the $\gamma$--ray flux, close to linearity, 
is likely linked to variations of the emissivity of the spine.
\end{abstract}
 

\section{Introduction}

The extragalactic sky at very--high energies (VHE; $E>50$ GeV) is dominated by blazars, 
radio--loud active galactic nuclei with relativistic jets pointing toward the Earth. 
This geometry is particularly favorable since, due to the relativistic aberration, the resulting 
non--thermal emission of the jet is strongly amplified and blue shifted. 
The spectral energy distribution (SED) of blazars displays two broad components, the one peaking in the IR--UV 
band produced by relativistic electrons through  synchrotron emission, and the high--energy one 
(peaking in $\gamma$--rays) attributed, in the so--called one--zone leptonic models,  
to the inverse Compton (IC) emission by the same electrons (e.g., Ghisellini et al. 1998).
The parameter regulating the amplification of the emitted radiation, the relativistic Doppler 
factor $\delta\equiv[\Gamma(1-\beta\cos\theta _{\rm v})]^{-1}$ (where $\Gamma$ is the bulk Lorentz 
factor of the flow and $\beta=v/c$), is strongly dependent on the angle between the jet axis and the line of sight, $\theta_{\rm v}$: 
it is maximal within a cone with semi--aperture $\theta _{\rm v}\simeq 1/\Gamma$ and drops rapidly outside it. 
In this scheme thus one expects that for viewing angles larger than $1/\Gamma$ the jet non--thermal 
luminosity becomes increasingly fainter. 
Accordingly to this expectation, only a handful of radiogalaxies -- thought to be the 
{\it parent} (i.e. not beamed at us) population of blazars --  have been detected in VHE $\gamma$--rays 
(Aharonian et al. 2006, 2009, Aleksic et al. 2012).

In the simple scheme sketched above, the SED of radiogalaxies should resemble that 
of blazars, the only difference being the reduced beaming. 
However, as discussed in e.g., Tavecchio \& Ghisellini (2008, TG08 hereafter) and Aleksic et 
al. (2014, A14 hereafter), the relatively large separation of the frequencies of two peaks, coupled to 
the expected low Doppler factor ($\delta\simeq 2-4$) is difficult to reproduce with one--zone emission models. 
Indeed, there are indications that the jet structure is not as simple as that assumed above. 
There is compelling observational evidence that the jet of low--power TeV blazars  
could be structured (e.g. Giroletti 2008), with a fast (with bulk Lorentz factor $\Gamma_{\rm s}=10-20$) 
central spine, surrounded by a slower ($\Gamma_{\rm l}=2-4$) layer. 
A structure of this type is also required to unify BL Lacs objects with the parent population of 
FRI radiogalaxies (Chiaberge et al. 2000). From 
the point of view of the radiative properties, such a structure leads to the enhancement of the 
radiative efficiency of both components, since the electrons of each region can inverse--Compton 
scatter the beamed soft photons coming from the other (Ghisellini et al. 2005). 
As shown in Ghisellini et al. (2005) this model can account for the VHE emission of TeV detected 
blazars (thought to be produced in the spine) and predicts that the weakly beamed emission of 
the layer could dominate the emission from misaligned radiogalaxies. 
This scheme was successfully applied to the modeling of the SED of the first radiogalaxy detected at VHE, M87 (TG08). 
The same idea is at the base of the  decelerating jet scenario, in which one assumes that the inner jet moves 
faster then the outer jet zones. 
Also in this scenario the faster and the slower regions can interact through their radiation fields 
(Georganopoulos \& Kazanas 2003).

NGC 1275, located in the Perseus cluster ($D\simeq 75$ Mpc) is one of the closest radiogalaxies. 
It has been the subject of intense study in the last years, particularly in view of the evident 
impact of the relativistic jets on the  gas of the surrounding cluster (e.g. Fabian et al. 2011). 
At high energy it was not detected by EGRET onboard {\it CGRO} flown in the nineties. 
Instead, {\it Fermi--}LAT detected NGC 1275 soon after its launch (Abdo et al. 2009), with a flux 4 times 
larger than the EGRET upper limit, implying secular variability of the high-energy flux. 
The $\gamma$--ray emission is variable also on much smaller timescales, down to a week (Kataoka et al. 2010, 
Brown \& Adams 2011). 
Radio observations revealed an increasing of the activity starting in 2005 (Abdo et al. 2009) 
and recent VLBI observations (Nagai et al. 2010, 2014) show a new radio component ejected from the core. 
NGC 1275 has been the third radiogalaxy detected in the VHE band (by MAGIC), showing a very soft 
spectrum smoothly connected to the GeV spectrum (Aleksic et al. 2012). 
A14 report multifrequency data obtained during two multifrequency campaigns held in 2010--2011, 
thanks to which the overall SED could be assembled. 
Unfortunately, the available X-ray spectrum, probing the critical region of the SED between 
the two bumps, is affected by strong pile-up and it was not possible to fix the slope of the underlying 
non--thermal power law. 
The SED is barely compatible with a one-zone leptonic model for the values of the viewing angle 
usually considered in literature. 

The problem with the one--zone model is that, with the small Doppler factors corresponding to the 
large $\theta_{\rm v}$, it is difficult to reproduce the required large separation of the 
frequencies of the two peaks. 
Similar to the case of M87, it is tempting to think that a solution is to admit a structured jet. 
In fact, strong observational support to this scenario is provided by the clear evidence for a 
spine--sheath structure shown in recent VLBA maps presented in Nagai et al. (2014). 
The observed limb--brightened structure suggests the presence of a faster central core and a 
slower sheath for the inner (i.e. parsec scale) jet. For an assumed angle of $25^{\rm o}$ 
and with the hypothesis of equal rest-frame emissivity, the data are consistent with a bulk 
Lorentz flow of the sheath of 2.4. 
Motivated by this observational evidence and by the difficulties of the one-zone model, in the following we will explore the applicability of the spine--layer scenario of Ghisellini et al. (2005) to the multifrequency emission of NGC 1275. 

After a discussion of the SED (\S2) we will describe the model and its application (\S3). 
We discuss the results in \S4. Throughout the paper, we assume the following cosmological parameters: 
$H_0=70$ km s$^{-1}$ Mpc$^{-1}$, $\Omega_{\rm M}=0.3$, $\Omega_{\Lambda}=0.7$.

\section{The spectral energy distribution}

\begin{figure}
\hspace{-3.5 truecm}
\psfig{file=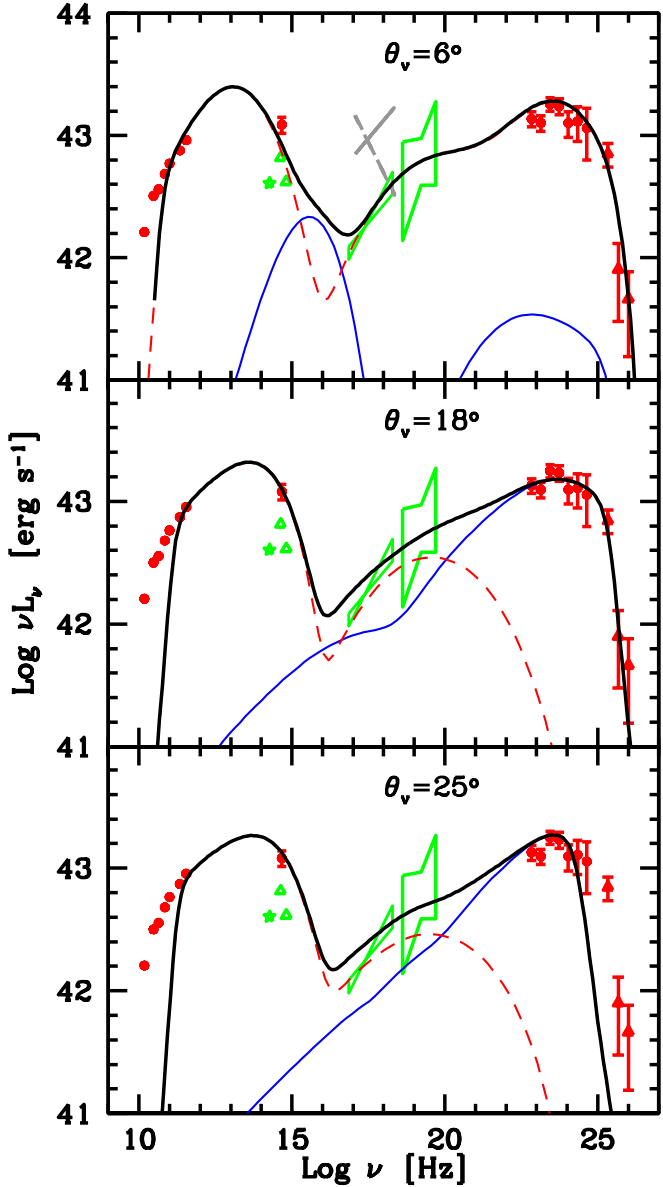,height=16.cm,width=16.cm}
\caption{
Spectral energy distribution of the core of NGC 1275 during the first MAGIC campaign 
reported in Aleksic et al. (2014). 
Red symbols show the quasi simultaneous multifrequency data. 
The green points in the optical band are historical data from Chiaberge et al. (1999) and 
Baldi et al. (2010). 
Green ``bow-ties" in the X--ray band are from Balmaverde et al. (2006) and Ajello et al. (2009). 
The dashed (solid) gray line in the top panel shows the {\it Chandra} spectrum assuming 
a photon index of 2.5 (1.7). 
We refer to Aleksic et al. (2014) for details. 
The three panels report the results of our structured jet emission model for two different values 
of the viewing angle, from top to bottom: 
$\theta_{\rm v}=6^{\rm o}$, $\theta_{\rm v}=18^{\rm o}$ and $\theta_{\rm v}=25^{\rm o}$. 
The red (dashed) and the blue  (solid) lines show the emission from the spine and the layer, respectively. 
The thick black line shows the total. 
See text for details.
}
\label{18deg}
\end{figure}

The SED of NGC 1275 during the first MAGIC campaign (A14) is reported in Fig. \ref{18deg}. 
Red symbols show almost simultaneous data, green lines are for historical data (see A14 for references). 
Although the covering of the low--frequency region is relatively sparse, one can identify two components, 
one traced by the radio and optical data, the other one well tracked by {\it Fermi}--LAT and MAGIC data points. 
The gray line shows a power law spectrum of the central emission from NGC 1275 derived from {\it Chandra} observations. 
As discussed in A14, the observation is affected by a strong pile--up. 
This, coupled with the complexity of the emission from the central regions of NGC 1275 (which includes 
several contributions, mainly the thermal emission of the cluster and host galaxy gas), 
makes impossible to determine a unique  fit. 
A14 report different fits of the X-ray spectrum, corresponding to different fixed values of the spectral 
index of the power law. 
Considering the goodness of the fit, the best solution would correspond to a relatively flat spectrum, 
with photon index 1.7. 
Other possible solutions, however, include steeper spectra. 
In particular A14 adopt the solution corresponding the a photon index of 2.5 (dashed gray line in the top panel of Fig. \ref{18deg}), 
which is compatible with the slope of the high--energy tail of the synchrotron component of the one--zone model. 
Although possible, this slope is however rather different than that previously derived for the soft and hard 
X--ray spectra (green bow--ties, Balmaverde et al. 2006, Ajello et al. 2009), which consistently trace 
a power law continuum with slope $\approx 1.7$. 
However, the flux corresponding to the latter slope (solid gray line in the left panel) is exceptionally high, 
not compatible with the optical datapoint and the LAT spectrum. 
Given these large uncertainties we choose to use the historical X-ray data (green) for our modeling. 
With this choice the low--energy components of the SED has to peak between the radio and the optical band, 
with the optical associated to the high--energy tail of the component. 
This is different from the model discussed in A14, for which the synchrotron component was instead located 
in the UV band, with the optical and the (steep) X--ray emission associated to the low and high 
frequency tails of the synchrotron emission.  

\section{Modelling}

\subsection{One--zone synchrotron-self Compton model}

As discussed A14, the modeling of the SED within the one-zone SSC scenario is challenging. 
A14 performed their analysis in the case of a large synchrotron frequency (in the soft X-ray band); 
their arguments are even more compelling in our interpretation, with the SED peak below the optical band. 
In fact, as detailed by TG08 for the case of M87, the main problem is represented by the large 
separation between the two SED peaks, which unavoidably implies huge Doppler factors, 
incompatible with the large viewing angle inferred for radiogalaxies. 
Adapting the limit obtained in TG08 for the numerical values suitable for NGC 1275, 
the SED constrains the Doppler factor to be larger than: 
\begin{equation}
\delta\simeq 450 \, L_{\rm s,43.5}^{1/2}\, L_{\rm IC,43.3}^{-1/4}\,  
\nu_{\rm s,13}^{-1} \, \nu_{\rm IC, 24}^{1/2} t_{\rm var, 1\; w}^{-1/2},
\end{equation}
where $L_{\rm s}$ and $L_{\rm IC}$ are the synchrotron and IC luminosities, 
$\nu_{\rm s}$ and $\nu_{\rm IC}$ the synchrotron and IC peak frequencies and we assume 
a (conservative) variability timescale $t_{\rm var}=1$ week.

We therefore exclude the one-zone SSC model and in the following we discuss the structured jet scenario.

\subsection{The structured jet scenario}

We briefly recall the main features and the parameters of the model fully described 
in Ghisellini et al. (2005) and TG08. 
The spine is assumed to be a cylinder with eighth $H_s$ and radius $R_s$. 
The subscripts ``s" and ``l" stand for spine and layer, respectively.
The layer is a hollow cylinder  with height  $H_l$, internal radius $R_s$ and 
external radius $R_l=1.2\times R_s$. 
As for  the case of M87, we assume that $H_s\ll H_l$, corresponding to the case 
of a perturbation (e.g. a shock) 
traveling down the the spine, surrounded by a relatively long and stationary layer 
of slow plasma, possibly resulting from the interaction of the edge of the jet with the external medium.

Electrons in both zones follow a broken power law energy distribution, specified by the minimum, 
the maximum and the break Lorentz factors  
$\gamma _{\rm min}$, $\gamma _{\rm max}$ and $\gamma _{\rm b}$ and indices $n_1$ and $n_2$. 
The model assumes that the electron energy distribution is stationary and it 
does not take into account the electron radiative losses. 
The normalization of the electron distribution is parametrized by the emitted 
synchrotron luminosity, $L_{\rm syn}$. 
The emitting regions are filled with a tangled magnetic field $B_{s}$, $B_{l}$. 
The relativistic beaming is specified by the two Lorentz factors $\Gamma _{s}$, $\Gamma_{l}$ 
and by the viewing angle $\theta_{\rm v}$.

Electrons emit  synchrotron and IC radiation. For 
the latter, besides the local synchrotron radiation field (SSC emission) we also consider 
the radiation field of the other component. 
Due to the relative motion, the energy density of the radiation field of one component in the 
rest frame of the other is boosted by the squared of the relative Lorentz factor, 
$\Gamma_{\rm rel}=\Gamma_{\rm s} \Gamma _{\rm l}(1-\beta_{\rm s}\beta_{\rm l})$.
In the calculations we also take into account the absorption of $\gamma$--rays through the 
interaction with the soft radiation field, $\gamma \gamma\rightarrow e^+ e^-$. 
Since the radiation is produced and absorbed within the same region, the ``suppression factor" due 
to absorption is $I(\nu)/I_{\rm o}(\nu)=\{1-\exp[-\tau_{\gamma \gamma}(\nu)]\}/\tau_{\gamma \gamma}(\nu)$ 
which, for large optical depths, simply becomes $1/\tau_{\gamma \gamma}$.

For a fixed viewing angle $\theta_{\rm v}$, the (bolometric) synchrotron and SSC luminosities 
of both components are amplified by the usual boosting factors 
$\delta_{\rm s}^4$ and $\delta_{\rm l}^4$. 
The IC emission from the scattering of the radiation field produced in the other component 
follows a more complex pattern, involving the relative Doppler factors between the two 
components (see TG08 for a detailed discussion).

Although the number of the parameters is large (almost twice that of the simple SSC model), 
the model has to satisfy other constraints that can be used as guidelines in selecting the suitable setup.

\subsubsection{Single emission region and external seed photons}

The first point that we consider is the possibility that the two peaks of the SED are produced 
by a single emitting region (hence by the same electron population) but with target photons for 
the IC scattering being provided by an external region (without this last condition we would 
fall in the SSC model previously excluded). 
In the structured jet scenario the emission region could be the layer and the external radiation 
is that produced by the spine, or {\it viceversa}. 
Indeed this was the original set-up discussed in Ghisellini et al. (2005), in which 
the entire SED of radio--galaxies was attributed to the layer.

This possibility is disfavored for the case of NGC 1275 (as for the previous case of M87 in TG08) 
because it requires a Doppler factor not compatible with the inferred relatively large 
viewing angle $\theta _{\rm v} \gtrsim 20^{\circ}$. Briefly, the reasoning leading to this 
conclusion develops through the following three steps : 
i) the frequency of the IC peak, $\nu_{\rm IC}$, provides a lower limit for the Lorentz factor 
of the electrons emitting at the peaks of the SED, $\gamma_{\rm p}$; 
ii) this limit, coupled to the observed synchrotron peak frequency $\nu_{\rm s}$,  
provides an upper limit for the ratio of the magnetic field and the Doppler factor $B/\delta$; 
iii) the observed high energy flux limits the level of the SSC component:
the magnetic field cannot be smaller than a critical value.
This, coupled with the $B/\delta$ upper limit derived above, constrains $\delta$ to be larger
than a critical value, that can be achieved only for small angles, no matter the value of $\Gamma$.

More quantitatively the steps are the following: 
\begin{itemize}

\item 
If the IC peak is produced in the KN regime, the energy of the electrons at the peak are 
comparable to the photons energies, i.e. 
$\gamma _{\rm p} m_{\rm e} c^2 \delta \simeq h\nu_{\rm IC}$. 
If, on the other hand, the peak is produced in the Thomson regime, the peak frequency would be 
$\nu_{\rm IC}\simeq \gamma^2_{\rm p}\nu^{\prime}_{\rm ext} \delta$ (where $\nu^{\prime}_{\rm ext}$ 
is the peak frequency of the external radiation field in the comoving frame of the emitting source) 
with the Thomson condition $\gamma _{\rm p}h \nu_{\rm ext}^{\prime}<m_{\rm e} c^2$, 
which provides $\gamma _{\rm p} m_{\rm e} c^2 \delta > h\nu_{\rm IC}$. 
Inserting the numerical values, the Lorentz factor of the electrons emitting at 
the peak would therefore be $\gamma _{\rm p}\gtrsim 8\times 10^{3} \delta^{-1} \nu_{\rm IC,24}$.

\item 
The observed synchrotron peak frequency 
$\nu_{\rm s}\simeq 2.8\times 10^{6} B\gamma _{\rm p}^2 \, \delta$ Hz 
with the limit derived above for $\gamma _{\rm p}$ implies $B/\delta< 0.056 \, \nu_{\rm s,13} \, 
\nu^{-2}_{\rm IC, 24}$ G.   

\item 
With the value of $\gamma _{\rm p}$ and $\nu_{\rm s}$, the SSC component is predicted to peak 
at a frequency $\nu_{\rm SSC}\simeq 10^{21}/\delta^2$ Hz. 
Its luminosity therefore is limited by the observed hard X-ray flux.
Looking at the SED we can write this condition as:
\begin{equation}
\frac{L_{\rm SSC}}{L_{\rm s}}=\frac{U^{\prime}_{\rm s}}{U_B} \lesssim 0.1 \; 
\to \; \frac{L_{\rm s}}{4\pi R^2 c \, U_B \delta^4}\lesssim0.1.
\end{equation}
Using the limit on $B$ derived above and expressing the radius as $R< c t_{\rm var} \delta$ we find: 
\begin{equation}
\delta\gtrsim 
3.5 \, L_{\rm s,43.5}^{1/8} \, \nu_{\rm IC,24}^{1/2} \, \nu_{\rm s,13}^{-1/4} 
t_{\rm var, 1w}^{-1/4},
\label{deltalimit}
\end{equation}
in which we have assumed the (conservative) value $t_{\rm var}=1$ week. 

\end{itemize}

Note that the case  $\gamma _{\rm p}= h\nu _{\rm IC}/m_{\rm e}c^2 \delta$ --- corresponding to 
scattering in the Thomson--KN transition --- is the most conservative one in terms Doppler factor. 
In fact, larger values of $\gamma _{\rm p}$ imply smaller magnetic fields and thus larger 
Doppler factors to keep the SSC luminosity below the limit. 
We remark that the above inference only
assumes that the seed photons for scattering are not synchrotron. 
Therefore the conclusion is valid not only for a structured jet, but also when
the external radiation originates in the environment surrounding the jet 
(e.g. from accretion or dust). 

The derived lower limit for the Doppler factor, although not extreme, 
requires $\theta _{\rm v}\lesssim 16.5^{\circ}$, 
smaller than the value $\theta _{\rm v}\gtrsim 20^{\circ}$
estimated by radio observations (see A14).
Moreover the situation is even more complex due to the fact that for 
$\theta_{\rm v}\sim 10^{\circ}$ and Lorentz factors $\Gamma_{\rm s}\sim 10$, $\Gamma_{\rm l}\sim 3$--4,
the Doppler factors of both components are  very similar, making unlikely
that a single component dominates the entire SED.
For smaller still
angles, the emission from the spine starts to dominate the SED. 
In Fig. \ref{18deg} (top panel) we report our fit for 
$\theta_{\rm v}=6^{\circ}$, implying that NGC 1275 is a blazar. 
The dashed red and the thin solid blue line show the contribution of the spine and the layer to the total 
emission, clearly dominated by the spine. 
The high-energy bump is a mix of SSC 
(in the X--ray band) and IC off the photons of the layer (the peak in the $\gamma$--ray band). 
The corresponding parameters are reported in the first raw of Tab. 1.

Although the model reproduces quite satisfactorily the SED, the implied angle is clearly 
not consistent with the values derived from radio observations. Given this difficulty, 
the possibility of a single emitting region seems rather unlikely. 
In the next section we therefore consider the case with a large 
$\theta_{\rm v}$ and with both regions contributing to the emission.

\begin{table*} 
\begin{center}
\begin{tabular}{ll|llllllllllll|}
\hline
\hline
&$R$    
& $H$  
& $L_{\rm syn}$  
& $B$  
& $\gamma_{\rm min} $  
& $\gamma_{\rm b} $ 
& $\gamma_{\rm max}$  
& $n_1$
& $n_2$
& $\Gamma$ 
& $\theta_{\rm v}$ \\
& cm  &cm &erg s$^{-1}$ & G & & &  & & & &deg. \\
\hline  
Layer & $5\times 10^{16}$ &$2\times 10^{16}$ &$10^{40}$  &0.3 &150 &$10^{4}$ &$2\times10^5$   &1   &3.3   &3 &6 \\ 
Spine & $5\times 10^{16}$ &$10^{15}$ &$2\times10^{40}$  &0.08 &60 &$1.9\times 10^3$ &$5\times 10^4$ &1.9 &3.5 &10  &6  \\
\hline  
Layer & $5\times 10^{16}$ &$2\times 10^{16}$ &$6.5\times10^{40}$  &0.9 &150 &150 &$10^6$   &1   &2.5   &4 &18 \\ 
Spine & $5\times 10^{16}$ &$10^{15}$ &$1.3\times10^{43}$  &2.5 &60 &$8\times 10^2$ &$1.5\times 10^4$ &2.3 &2.5 &10  &18  \\
\hline
Layer & $5\times 10^{16}$ &$2\times 10^{16}$ &$5\times10^{41}$  &2 &600 &100 &$5\times10^6$   &1   &2.5   &4 &25 \\ 
Spine & $5\times 10^{16}$ &$10^{15}$ &$1.5\times10^{44}$  &8.5 &5 &$ 5$ &$1.5\times 10^4$ &1 &2.5 &10  &25  \\
\hline
\hline
\end{tabular}                                                         
\caption{Input parameters of the models for the layer and the spine
shown in Fig. \ref{18deg}. All quantities (except the bulk Lorentz
factors $\Gamma$ and the viewing angle $\theta_{\rm v}$) are measured
in the rest frame of the emitting plasma.  The external radius of the
layer is fixed to the value $R_2=1.2 \times R$. }
\end{center}
\label{tab1}
\end{table*}                                                                  

\subsubsection{Two--component model}

After the discussion in the previous section we are left with two alternative possibilities: 
either the spine produces the observed synchrotron component and the layer the high--energy component or, 
conversely, the spine produces the $\gamma$--ray  emission and the layer the low frequency peak. 

In the latter case, the spine would produce a SED strongly unbalanced toward the IC peak, since the 
synchrotron luminosity must be low enough not to contribute to the observed emission, $L_{\rm IC}\gg L_{\rm s}$. 
In this case, the corresponding emission from the spine as seen by an observer aligned with the jet 
would be even more extreme, since the ratio between the IC (by construction mainly produced by the 
scattering of the photons from the layer) and the synchrotron luminosities increases for smaller angles 
(Dermer 1995, TG08). 
The resulting source would appear as a strongly IC dominated BL Lac object, for which there 
are no representatives in current samples of blazars. 

The most plausible solution is thus to associate the low--energy component with the synchrotron 
emission of the spine and the high--energy bump to the IC emission of the layer. 
This is the solution we discuss in the following.


In Fig. \ref{18deg} (second and third panels) we report the results of our model 
for two different viewing angles, $\theta_{\rm v}=18^{\rm o}$ and $25^{\rm o}$. 
As discussed above, we suppose that the spine (red line) is responsible for the low energy bump, 
while the high-energy component is the result of the layer IC component (blue). 
To produce $\gamma$--rays, the electrons of the layer have quite large Lorentz factors, 
$\gamma_{\rm max}\sim10^6$. 
The corresponding synchrotron frequency falls in the X--ray band. 
To not overproduce the flux in this band we keep a low value of the layer magnetic field. 
Clearly, the emission from the layer is strongly unbalanced in favor of the IC emission, 
dominated by the scattering off the IR photons of the spine. 
When setting the parameters of the spine we are guided by the requirements to produce a 
SED dominated by the synchrotron component. 
However, we also require that the IC component is not too low. 
This requirement follows from the fact that, once observed at smaller angles, one expects to 
observe a SED similar to that of typical low synchrotron peaked BL Lac objects (see below). 
In this case it is likely that the IC component, peaking in the soft $\gamma$--ray band, substantially 
contributes to the observed X--ray continuum. 
Note that, due to the low luminosity and the relatively high frequency of the synchrotron peak of 
the layer, the resulting contribution of this radiation field to the IC emission from the spine 
is negligible (thus the IC peak is dominated by the SSC component). 
In analogy with the case of M87, we adopt bulk Lorentz factors $\Gamma_{\rm s}=10$ and $\Gamma_{\rm l}=4$. 
The relative Lorentz factor is $\Gamma_{\rm rel}=1.46$. 
The value of the spine is typically derived for BL Lac objects (e.g. Ghisellini et al. 2010), 
sources for which the spine should dominate the entire SED. 
In the case of the layer, we adopt a value similar to that used for M87. 
Having fixed these parameters we try to find the set of the remaining parameters that better reproduce the data.

\begin{figure}
\hspace{-.3 truecm}
\psfig{file=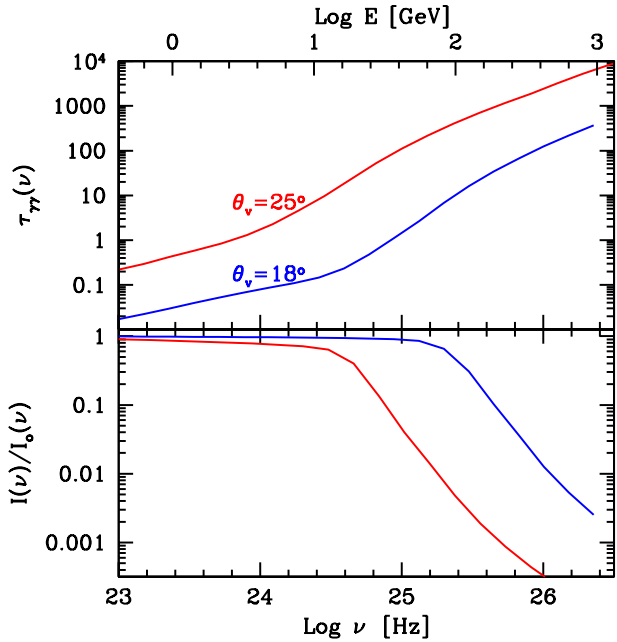,height=9.cm,width=9.cm}
\vspace{-.9 truecm}
\caption{Optical depth (upper panel) and suppression factor (lower panel) 
for absorption of $\gamma$--rays within the jet as a function of the frequency for the two models 
reported in Fig. \ref{18deg}, for $\theta_{\rm v}=18^{\rm o}$ (blue) and $\theta_{\rm v}=25^{\rm o}$ (red).
}
\label{tau}
\end{figure}

For $\theta_{\rm v}=18^{\circ}$ the model can satisfactorily reproduce the observed high--energy bump. For 
$\theta_{\rm v}=25^{\circ}$, however, the model fails in reproducing the data at the highest energies. 
The reason for this problem, already discussed in TG08 for the case of M87 but more severe here, 
is the huge photon--photon absorption of the high-energy $\gamma$--rays produced in the 
layer by the dense IR--optical radiation field of the spine. 
While for the case $\theta_{\rm v}=18^{\rm o}$ the intrinsic (i.e. beaming corrected) synchrotron luminosity 
of the spine is still low enough to make the source transparent for photons at few hundreds of GeV, 
for $\theta_{\rm v}=25^{\rm o}$ the optical depth exceeds unity  above few tens of GeV, 
causing an abrupt cut--off of the emission. 
The optical depth can be approximated by (e.g. Dondi \& Ghisellini 1995):
\begin{equation}
\tau_{\gamma \gamma}(E)\simeq \frac{\sigma _{\rm T}}{5} r \, n^{\prime}(\epsilon^{\prime}) \epsilon^{\prime}
\label{taueq}
\end{equation}
where $r$ is the jet radius and $n^{\prime}(\epsilon^{\prime}) \epsilon^{\prime}$ is the soft photons 
number density in the layer frame, calculated at the threshold energy $\epsilon^{\prime}=m^2c^4\delta/E$:
\begin{equation}
n^{\prime}(\epsilon^{\prime}) \epsilon^{\prime}\simeq \frac{\epsilon^{\prime} 
L^{\prime}_s(\epsilon^{\prime})}{4\pi r^2 c \, \epsilon^{\prime}}= 
\frac{\epsilon L_s(\epsilon) \Gamma^2_{\rm rel}}{4\pi r^2 c \, \epsilon^{\prime} \delta_{\rm s}^4}
\end{equation}
where $\delta^4_{\rm s}$  is used to transform the observed synchrotron luminosity to the spine rest frame 
and  $\Gamma^2_{\rm rel}$  is the boosting term of the photon energy density from the spine to the layer frame.
For a fixed observed luminosity of the target photons,  the optical depth depends strongly on the Doppler 
factor, as $\delta_{\rm s}^{-4}$. 
In Fig. \ref{tau}, we  show the optical depth and the corresponding ``suppression factor" 
$[1-\exp(-\tau_{\gamma \gamma})]/\tau_{\gamma \gamma}$ as a function of the frequency, in the two cases. 
The shape of the curves, which reflects the spectrum of the target photons, is similar in both cases, 
but the case $\theta_{\rm v}=25^{\rm o}$ implies an optical depth larger by a factor of $\approx10$, 
as expected from the ratio of the two boosting factors in the two cases, $(1.86/1.02)^4\sim11$. For 
$\theta_{\rm v}=18^{\rm o}$ the source is transparent ($\tau_{\gamma \gamma}<1$) up to $\sim 100$ GeV. 
Instead, for $\theta_{\rm v}=25^{\rm o}$, the optical depth reaches unity already at 
20--30 GeV and rapidly increases, determining the abrupt cut--off visible in Fig. \ref{18deg} (lower panel).

An effect possibly mitigating the importance of the opacity, not treated in our calculations, 
is related to the anisotropy of the soft photon target field. 
In fact, in the frame of the layer, 
the $\gamma$--ray photons eventually reaching us make an angle of $\sim$90$^\circ$
with the jet axis, while the target photons, produced by the spine, are going along the 
jet axis.
The $\gamma$--rays and their targets therefore preferentially collide with an angle of $90^\circ$.

Since the threshold condition for the reaction depends on the angle, 
$\epsilon^{\prime}E^{\prime}=2m^2c^4/(1-\cos\theta)$, this translates into a shift to 
higher $\nu$ by a factor $\approx 2$ of the curves in Fig. \ref{tau}.
This, coupled to  the presence of another factor $(1-\cos\theta)$ in the expression of the optical depth, 
implies an effective decrease of $\tau$ at a given frequency. 
However, while the effect could be important  
for frequencies for which $\tau\sim 1$ (i.e. around $\nu=10^{24}$ Hz for the case 
$\theta_{\rm v}$=25$^{\circ}$), its impact  on the suppression factor at larger frequencies, 
for which the optical depth rapidly reaches $\tau\gg 1$, is almost negligible.

Focusing on the case for $\theta_{\rm v}=18^{\rm o}$, the power carried by the jet 
(assuming a composition of one cold proton per emitting electron, e.g. Celotti \& Ghisellini 2008) is 
$P_{\rm j,l}=6\times 10^{43}$ erg s$^{-1}$ and $P_{\rm j,s}=2.3\times 10^{47}$ erg s$^{-1}$ for the 
layer and the spine, respectively. 
The corresponding radiative luminosities (corrected for beaming) are $
L_{\rm r,l}=2\times 10^{43}$ erg s$^{-1}$ and $L_{\rm r,s}=2\times 10^{45}$ erg s$^{-1}$. 
The coupling with the spine radiation field makes the layer extremely efficient 
to convert kinetic power into radiation, with an efficiency $\eta_{r,l}=P_{\rm j,l}/L_{\rm r,l}\simeq30\%$. 
The spine power appears quite large, more similar to the average power inferred for the 
powerful flat spectrum radio quasar than those of BL Lac objects (e.g., Ghisellini et al. 2010). 
The mass of the black hole in NGC 1275 is estimated to be around $M_{\rm BH}\sim3\times 10^8 M_{\odot}$ 
(Woo \& Urry 2002), corresponding to an Eddington luminosity of $L_{\rm Edd}\approx 4.2\times 10^{46}$ 
erg s$^{-1}$. The jet power is thus about a factor of 5 larger, consistent with what generally 
inferred by studies of the jet of blazars (e.g. Ghisellini et al. 2010).
 
\begin{figure}
\hspace{-.5 truecm}
\psfig{file=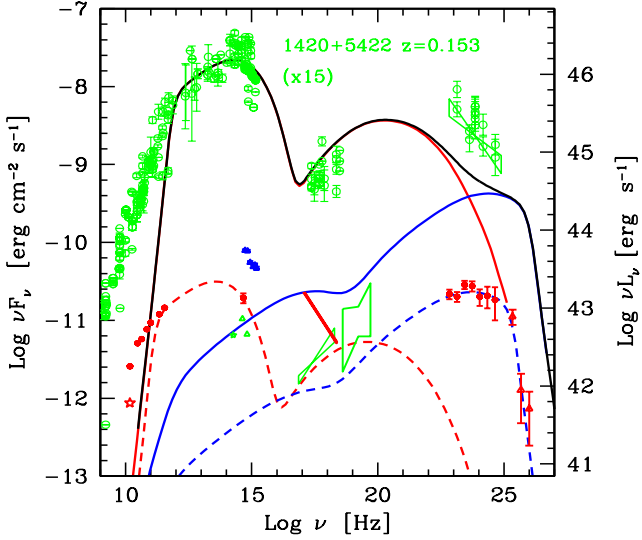,height=8.5cm,width=9.cm}
\vspace{-.9 truecm}
\caption{The black line shows the emission from the spine-layer model of NGC 1275 for 
the case $\theta_{\rm v}=18^{\circ}$ (left panel in Fig. 1 and dashed red and blue lines)
as observed at $\theta_{\rm v}=6^{\circ}$. 
The solid red and blue lines shows the separate contribution of the spine and the layer, respectively. For 
comparison, the green data points (associated to the luminosity, right $y$-axis) describe the 
SED of 1420+5422, a BL Lac at $z=0.153$ with a luminosity multiplied by a factor of 15 
(data from {\tt http://tools.asdc.asi.it}).
}
\label{1420}
\end{figure}

Fig. \ref{1420} shows the SED calculated for the case $\theta_{\rm v}=18^{\circ}$  as would be recorded 
by an observer located at $\theta_{\rm v}=6^{\circ}$ from the jet axis (solid black line). 
This SED would thus correspond to that of the blazar associated to the misaligned jet of NGC 1275 
(numerical values of the flux correspond to the blazar located at the same redshift as NGC 1275). 
The synchrotron peak located in the optical band would lead to classify the source as a BL Lac object 
with the peak in the IR band. 
The overall shape of the SED observed at $\theta_{\rm v}=6^\circ$ 
resembles (apart from the $\gamma$--ray emission) 
the shape of the BL Lac object 1420+5422 ($z=0.153$, Shaw et al. 2013),
which is however a factor $\sim$15 less luminous than the ``aligned" NGC 1275.
In fact the data of 1420+5422 (green data points, 
from {\tt http://tools.asdc.asi.it}) have been increased by a factor 15.


Finally we discuss the possible role of an environmental radiation field. 
In fact, any soft ambient radiation field, beamed in the layer frame, could contribute to the IC emission. 
Unfortunately the knowledge of the nuclear environment of NGC 1275 is far from complete. 
As for the other FRI radio galaxies, the lack of broad emission lines leads to the idea 
that any accretion flow should be quite inefficient. 
We can derive a robust upper limit on the luminosity of the nuclear component from the measured 
IR flux of the core (green star in Fig. \ref{18deg}), likely associated to the non--thermal emission from 
the inner regions of the jet (Baldi et al. 2010), $L_{\rm env}<3\times ~10^{42}$ erg s$^{-1}$.
The contribution of the external radiation is maximized if its emission occurs within a region with 
size of the order of the distance of the emission region in the jet, $d$. 
Larger sizes would lead to the dilution of the energy density (proportional to $1/d^2$), while for 
lower distances the radiation field in the jet frame would be deebosted. 
Assuming a jet aperture angle $\psi$, the distance of the emission region is $d\sim r/\psi$, and the ratio 
of the external energy density to the spine energy density in the frame of the layer can be estimated to be:
\begin{equation}
\frac{U^{\prime}_{\rm env}}{U^{\prime}_{\rm s}}< \frac{L_{\rm env}}{L_{\rm s}} 
\left( \frac{\Gamma_{\rm l}}{\Gamma_{\rm rel}}\right)^2 \psi^{2} \delta^4_{\rm s}.
\end{equation}
Inserting the numerical values (assuming $\psi=0.1$) we find that $U^{\prime}_{\rm env}/U^{\prime}_{\rm s}<0.1$ 
and thus any possible radiation field can be safely neglected. 
Besides being a source of target photons for the IC process, an external radiation field 
could also act as an absorber for the $\gamma$--rays  produced by the jet. For 
an external radiation field with a narrow spectrum peaking at an energy 
$\epsilon_{\rm ext}=h\nu_{\rm ext}=0.4 \, \nu_{\rm ext,14}$ eV, 
the maximum of the optical depth would correspond to an energy 
$E=m^2c^4/\epsilon_{\rm ext}=0.6 \, \nu_{\rm ext,14}^{-1}$ TeV, 
i.e close to the MAGIC points at the highest energies. 
A simple estimate of the optical depth at this energy yields $\tau_{\gamma\gamma}\lesssim 2/d_{18}$ 
(assuming a distance $d\approx10^{18}$ cm).
This upper limit to the optical depth due to the environment is then smaller than the internal
absorption.


\section{Discussion}

Summarizing, we have shown that the overall SED of NGC 1275 can be satisfactorily 
reproduced in the framework of the ``spine-layer" model --- with bulk Lorentz factors 
for the two components similar to those used in the case of M87 --- if the viewing angle is smaller 
then $\theta_{\rm v}\sim 20$ deg.
Larger angles inevitably lead to a drastic suppression of the emission in the MAGIC energy band caused by the strong internal absorption of the $\gamma$--rays with energies above few tens of GeV, determined by the 
luminous IR radiation field associated to the spine emission. 

Estimates of the angle at pc scale, based on the detection of the counter jet, provide values  
in the range 30--55 degrees (Walker et al. 1994, Asada et al. 2006). 
However, smaller angles are often assumed (e.g. Abdo et al. 2009). 
A much lower value ($\theta_{\rm v}\approx 3^{\rm o}$) at sub--pc scale was derived from VLBI 
observations by Krichbaum et al. (1992). 
Considering these uncertainties and the not unlikely possibility that the jet bends from the 
sub--pc scale (where we suppose that the emission occurs) to the pc scales imaged with VLBI, 
we conclude that the spine--layer scenario is clearly constrained but still barely suitable 
to reproduce the data. 
A strong test able to seriously threaten the model would be the detection of photons at energies above 
1 TeV, for which, even with $\theta_{\rm v}=18^{\rm o}$, the  optical depth $\tau_{\gamma \gamma}>10^3$, 
corresponding to a suppression of the flux exceeding $10^3$.

A point not completely satisfactory about our preferred model for $\theta_{\rm v}=18^{\rm o}$ 
concerns the emission the jet would present if observed at smaller angles. 
As previously discussed in TG08, in the unification scheme for radio--galaxies and blazars 
we generally expect that, once observed at small angles, the non--thermal continuum from a 
radiogalaxy resembles that of a blazar. In particular, for NGC 1275 we expect that the SED 
shape follows that of high--power BL Lac objects. 
We indeed found a source 
whose SED shape closely traces our theoretical curve, but with a luminosity 
about one order of magnitude smaller than the beamed emission of NGC 1275.

A14 report the existence of a clear correlation between the LAT $\gamma$--ray flux, $F_{\gamma}$, 
and the optical flux, $F_{\rm opt}$, compatible with both a linear 
($F_{\gamma}\propto F_{\rm opt}$) and a quadratic ($F_{\gamma}\propto F_{\rm opt}^2$) dependence. 
In our scheme the flux in the two bands is dominated by the emission produced in two separate regions, 
i.e. the spine for the optical and the layer for the $\gamma$--rays. 
In principle several possibilities are allowed by this configuration. For 
instance, $\gamma$--ray variability without counterpart in the optical is possible with 
changes of the parameters of the layer (in particular those specifying the electron energy distribution) 
and a stationary spine emission. 
On the other hand, since the IC emission of the layer --- responsible for the observed high--energy emission ---
is dominated by the scattering of the synchrotron photons of the spine, we expect that variations 
of the optical emission are accompanied by variations in the $\gamma$--ray band. 
If such variations are driven by changes in the intrinsic emissivity of the spine, without 
changes in the structural parameters (in particular the bulk Lorentz factor) we expect a linear correlation. 
However, if variations of $\Gamma_{\rm s}$ (and/or of $\Gamma_{\rm l}$) are also involved, more 
complex patterns are possible, connected to the interplay between the beaming of the emission 
in the observer frame and the amplification of the energy density of the seed photons, depending 
on the relative motion of the spine and the layer. 
With such a complex phenomenology is difficult to clearly identify a mechanism for the observed correlation. 
The simplest situation, corresponding to a linear correlation, 
is associated to variations
of the electron energy distribution or of the magnetic field of the spine.

\section*{Acknowledgments}
We thank the referee for the insightful and constructive comments that helped us to 
greatly improve the paper. FT acknowledges  contribution from a grant PRIN--INAF--2011. 
Part of this work is based on archival data, software or on-line services provided by 
the ASI Science Data Center.


\begin{thebibliography}{}

\bibitem[\protect\citeauthoryear{Abdo et al.}{2009}]{2009ApJ...699...31A} Abdo A.~A., et al., 2009, ApJ, 699, 31 

\bibitem[\protect\citeauthoryear{Ajello et al.}{2009}]{2009ApJ...690..367A} Ajello M., et al., 2009, ApJ, 690, 367 

\bibitem[\protect\citeauthoryear{Aharonian et al.}{2006}]{2006Sci...314.1424A} Aharonian F., et al., 2006, Sci, 314, 1424 

\bibitem[\protect\citeauthoryear{Aharonian et al.}{2009}]{2009ApJ...695L..40A} Aharonian F., et al., 2009, ApJ, 695, L40 

\bibitem[\protect\citeauthoryear{Aleksi{\'c} et al.}{2012}]{2012A&A...539L...2A} Aleksi{\'c} J., et al., 2012, A\&A, 539, L2 

\bibitem[\protect\citeauthoryear{Aleksi{\'c} et al.}{2014}]{2014A&A...564A...5A} Aleksi{\'c} J., et al., 2014, A\&A, 564, A5 

\bibitem[\protect\citeauthoryear{Asada et al.}{2006}]{2006PASJ...58..261A} Asada K., Kameno S., Shen Z.-Q., Horiuchi S., Gabuzda D.~C., Inoue M., 2006, PASJ, 58, 261 


\bibitem[\protect\citeauthoryear{Baldi et al.}{2010}]{2010ApJ...725.2426B} Baldi R.~D., et al., 2010, ApJ, 725, 2426 

\bibitem[\protect\citeauthoryear{Balmaverde, Capetti, \& Grandi}{2006}]{2006A&A...451...35B} Balmaverde B., Capetti A., Grandi P., 2006, A\&A, 451, 35 

\bibitem[\protect\citeauthoryear{Brown \& Adams}{2011}]{2011MNRAS.413.2785B} Brown A.~M., Adams J., 2011, MNRAS, 413, 2785 

\bibitem[\protect\citeauthoryear{Celotti \& Ghisellini}{2008}]{2008MNRAS.385..283C} Celotti A., Ghisellini G., 2008, MNRAS, 385, 283 

\bibitem[\protect\citeauthoryear{Chiaberge, Capetti, \& Celotti}{1999}]{1999A&A...349...77C} Chiaberge M., Capetti A., Celotti A., 1999, A\&A, 349, 77 

\bibitem[\protect\citeauthoryear{Chiaberge et al.}{2000}]{2000A&A...358..104C} Chiaberge M., Celotti A., Capetti A., Ghisellini G., 2000, A\&A, 358, 104 

\bibitem[\protect\citeauthoryear{Dondi \& Ghisellini}{1995}]{1995MNRAS.273..583D} Dondi L., Ghisellini G., 1995, MNRAS, 273, 583 

\bibitem[\protect\citeauthoryear{Fabian et al.}{2011}]{2011MNRAS.418.2154F} Fabian A.~C., et al., 2011, MNRAS, 418, 2154 

\bibitem[\protect\citeauthoryear{Georganopoulos \& Kazanas}{2003}]{2003ApJ...594L..27G} Georganopoulos M., Kazanas D., 2003, ApJ, 594, L27 

\bibitem[\protect\citeauthoryear{Ghisellini et al.}{1998}]{1998MNRAS.301..451G} Ghisellini G., Celotti A., Fossati G., Maraschi L., Comastri A., 1998, MNRAS, 301, 451 

\bibitem[\protect\citeauthoryear{Ghisellini, Tavecchio, \& Chiaberge}{2005}]{2005A&A...432..401G} Ghisellini G., Tavecchio F., Chiaberge M., 2005, A\&A, 432, 401 

\bibitem[\protect\citeauthoryear{Ghisellini et al.}{2010}]{2010MNRAS.402..497G} Ghisellini G., Tavecchio F., Foschini L.,  Ghirlanda G., Maraschi L., Celotti A., 2010, MNRAS, 402, 497 


\bibitem[\protect\citeauthoryear{Giroletti et al.}{2008}]{2008A&A...488..905G} Giroletti M., Giovannini G., Cotton W.~D., Taylor G.~B., P{\'e}rez-Torres M.~A., Chiaberge M., Edwards P.~G., 2008, A\&A, 488, 905 

\bibitem[\protect\citeauthoryear{Kataoka et al.}{2010}]{2010ApJ...715..554K} Kataoka J., et al., 2010, ApJ, 715, 554 

\bibitem[\protect\citeauthoryear{Krichbaum et al.}{1992}]{1992A&A...260...33K} Krichbaum T.~P., et al., 1992, A\&A, 260, 33 

\bibitem[\protect\citeauthoryear{Nagai et al.}{2010}]{2010PASJ...62L..11N} Nagai H., et al., 2010, PASJ, 62, L11 

\bibitem[\protect\citeauthoryear{Nagai et al.}{2014}]{2014ApJ...785...53N} Nagai H., et al., 2014, ApJ, 785, 53 

\bibitem[\protect\citeauthoryear{Shaw et al.}{2013}]{2013ApJ...764..135S} Shaw M.S., et al., 2013, ApJ, 764, 135

\bibitem[\protect\citeauthoryear{Tavecchio \& Ghisellini}{2008}]{2008MNRAS.385L..98T} Tavecchio F., Ghisellini G., 2008, MNRAS, 385, L98 

\bibitem[\protect\citeauthoryear{Walker, Romney, \& Benson}{1994}]{1994ApJ...430L..45W} Walker R.~C., Romney J.~D., Benson J.~M., 1994, ApJ, 430, L45 

\bibitem[\protect\citeauthoryear{Woo \& Urry}{2002}]{2002ApJ...579..530W} Woo J.-H., Urry C.~M., 2002, ApJ, 579, 530 


\end{thebibliography}
\end{document}